# Fabrication of Bi2212 Cross Whiskers Junction

Yoshihiko Takano [a,b], Takeshi Hatano [a,b], Akira Ishii [a], Akihiro Fukuyo [c], Yoshimasa Sato [d], Shunichi Arisawa [a,b], and Kazumasa Togano [a,b]

[a] National Research Institute for Metals, Sengen, Tsukuba 305-0047 Japan
(National Institute for Materials Science)
[b] CREST, Japan Science and Technology Corporation, Japan
[c] Fac. Sci&Tech, Science Univ. of Tokyo, Yamazaki, Noda 278-8510 Japan
[d] Material Science, Univ. of Tsukuba, Tsukuba Japan

An intrinsic Josephson junction has been successfully fabricated without any micro-fabrication technique. Two Bi2212 whiskers were crossed with one another and joined by post-annealing. The inter-whisker electrical transport properties were measured by the four-probe method. The temperature dependence of resistance exhibited metallic behavior above $T_C$. The resistance decreased to zero around 80K, corresponding to the superconducting transition. The current-voltage characteristics at 5K exhibited a small hysteresis and voltage jump, which can be explained by the intrinsic Josephson effect.

## Introduction

The intrinsic Josephson effect observed in the layered crystal structure of high-$Tc$ cuprate superconductors has received much attention because of future perspective of useful devices [1-2]. The fabrication of the an intrinsic Josephson junction commonly involves advanced micro-fabrication technique, such as focused-ion-beam (FIB) or high-resolution photolithography [3-4]. For further investigation, a simple fabrication technique for the intrinsic Josephson junction is required. Here, we devised a cross whiskers junction as a new fabrication method for an intrinsic Josephson junction without micro-fabrication. The details of the fabrication process and transport properties of Bi2212 cross whiskers junction are presented.

## Experimental

Bi2212 whiskers were prepared from the following starting materials: $Bi_2O_3$, $SrCO_3$, $CaCO_3$ and $CuO$ [5-7]. A mixture of powders with the ratio of Bi:Sr:Ca:Cu=3:2:2:4 was melted at 1100-1200°C for 30min and subsequently quenched. The obtained amorphous samples were calcined at around 850 °C for 5 days in flowing 72% $O_2$-$N_2$ gas [8].

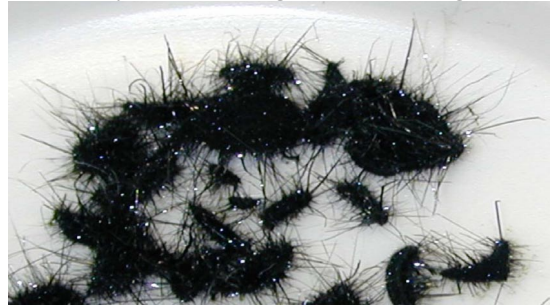

Fig.1
Photograph of Bi2212 whisker samples.

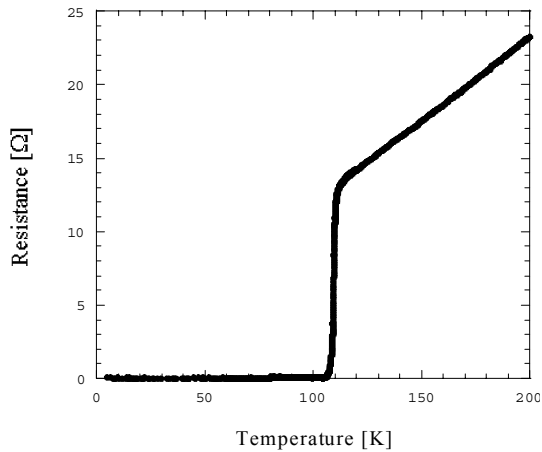

Fig 2.
  Temperature dependence of resistance of a whisker. Current level was 0.01mA along a-axis.

The obtained whiskers have a shape of ribbon with c-axis perpendicular to the surface and a-axis parallel to the longitudinal direction. The structure of the whiskers were examined by X-ray diffraction. Transport properties were measured by the standard four-probe method.

A cross whiskers junction was fabricated on an MgO(100) single-crystal substrate. MgO was chosen because the thermal expansion of MgO is comparable to that of a Bi2212 superconductor. Two whiskers were crossed perpendicularly on an MgO substrate. The c plane of the whiskers was parallel to the substrate surface. Subsequently, post-annealing was carried out at 850°C for 30 min. For resistance measurement, using Ag paste, Au lead wires were attached to both ends of the whiskers as four terminals.

### Results and Discussion

Figure 1 shows the as-grown whisker samples on an alumina crucible. The X-ray diffraction pattern showed that the whisker consists of only a Bi2212 phase, and the ratio of the intergrowth of 2223 phase was estimated to be less than 1% by lattice constant analysis [8]. Figure 2 displays the temperature dependence of resistance of a whisker. The current direction is parallel to the a-axis. With decreasing temperature, the resistance dropped to the order of 4 at 110K and exhibited a complete zero around 80K. An anomaly at 110K suggests the existence of an intergrowth of the 2223 phase.

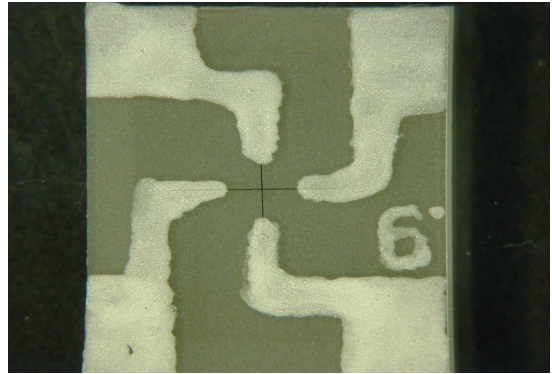

Fig 3.
  Photograph of Bi2212 cross whiskers junction fabricated on MgO substrate.

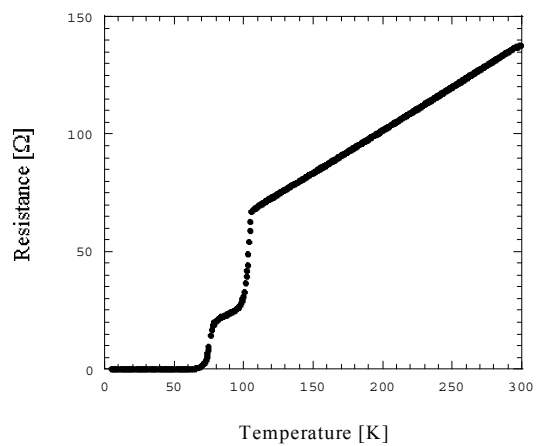

Fig. 4
  Temperature dependence of resistance for cross whiskers junction.

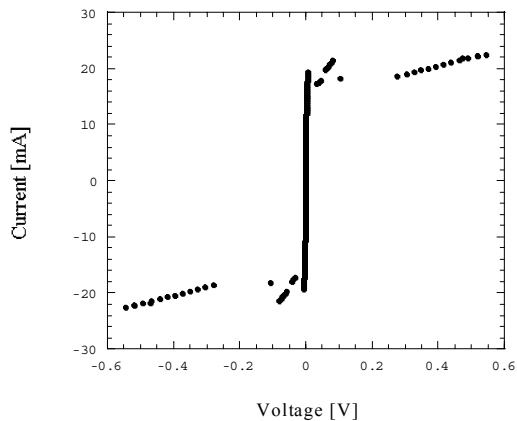

Fig 5(a).

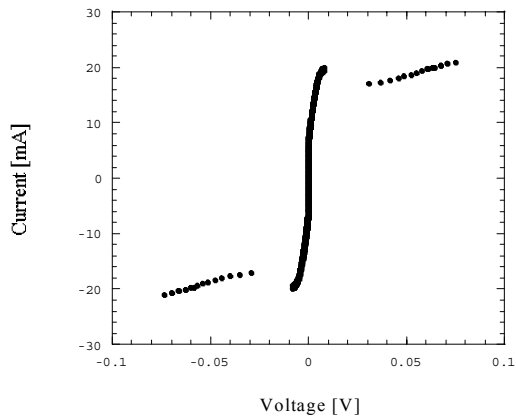

Fig 5(b).
(a) I-V characteristics of cross whiskers junction. (b) in the expanded scale.

A photograph of the cross whiskers junction is shown in Fig. 3. The widths of the whiskers were approximately 35 and 40μm, and the area of junction was estimated to be 1400μm². The temperature dependence of resistance of the inter-whisker is shown in Fig. 4. At high temperatures, metallic behavior was observed. With decreasing temperature, prior to the zero resistance at 70K, the temperature vs. resistance curve again exhibited an anomaly around 105K, corresponding to the change of the current path because of the superconducting transition of the Bi2223 intergrowth. Figure 5 (a) shows the current-voltage characteristics of the cross whiskers junction at 5K, and the expanded scale is plotted in Fig 5 (b). A clear hysteresis and voltage jump were observed. The magnitude of the first voltage jump was approximately 20mV, in accordance with one Josephson junction [10]. The critical current density $J_C$ was estimated to be 1400 A/cm².
## Summary

We have successfully fabricated a Bi2212 cross whiskers junction as an intrinsic Josephson junction. The cross whiskers junction can be easily fabricated by only one furnace. We are now doing further investigation of improving the processing parameters of whiskers and junctions. A novel and simple fabrication process of intrinsic Josephson junction will make an important role in the development of high-$T_c$ devices.